**Application of the Lagrange-Souriau form method to the case of source-free electromagnetic field in a curved space-time**

S.I. Tertychniy *

Institute for Theoretical Physics
University
D-50923 Cologne
Germany

**Abstract**

The recent method of the description of classical fields in terms of Lagrange-Souriau form is applied to the case of source-free electromagnetic field in order to check its computational capabilities. The relevant calculations are represented in all details and yield a useful data for comparisons of the method with more usual approaches in this simple and transparent case.

In the recent papers a new general method of description of systems of classical particles and fields has been suggested [1,2]. It seems interesting to see, however, its advantages (or perhaps disadvantages) in applications to concrete problems, at the first time, well known and easily tractable by means of more usual methods.

As one of the simplest examples, we shall consider a source-free electromagnetic (E-M) field on a space-time that is admitted to be curved. The inverse influence of E-M field to the space-time is neglected and, thus, the metric is assumed to be known. Let us denote space-time coordinates as $x^\mu$, electromagnetic potential as

$$\boldsymbol{\mathcal{A}} = \psi_\alpha dx^\alpha .$$

In the formalism under consideration *field variables* $\psi_\alpha$ are interpreted as quite independent coordinates (together with $x^\mu$) of some multi-dimensional *evolution space*. Moreover, all the derivatives of field variables entering the ordinary Lagrange function describing the field are considered as independent coordinates as well. In our case one has the additional coordinates $\chi_{\alpha\mu}$ that correspond to to partial derivatives of potentials $\frac{\partial\psi_\alpha}{\partial x^\mu}$. It is useful to introduce the following notations here:

$$\delta\psi_\alpha \equiv d\psi_\alpha - \chi_{\alpha\mu} dx^\mu, \qquad \frac{\delta}{\delta x^\mu} \equiv \frac{\partial}{\partial x^\mu} + \chi_{\alpha\mu}\frac{\partial}{\partial \psi_\alpha}; \qquad \Rightarrow \qquad \frac{\delta}{\delta x^\mu} \rfloor\, \delta\psi_\alpha = 0.$$

* On leave of absence from: The National Research Institute For Physical Technical and Radio-Technical Measurements (VNIIFTRI), Mendeleevo, Moscow Region, 141570, Russia

In frames of formalism under consideration the physical field is described by the closed Lagrange-Souriau p-form $\sigma$ on the space with coordinates $\{x^\mu, \psi_\alpha, \chi_{\alpha\mu}\}$. $\sigma$ may be locally represented as exact differential

$$\sigma = d\theta \quad \text{(locally)},$$

of the generalized Cartan form $\theta$ in turn defined by the equation

$$\theta = \epsilon_{\mu_1\cdots\mu_n} \sum_{k=0}^{n} \frac{1}{k!} C_n^k L^{\mu_1\cdots\mu_k|\alpha_1\cdots\alpha_k} \delta\psi_{\alpha_1} \wedge \cdots \wedge \delta\psi_{\alpha_k} \wedge dx^{\mu_{k+1}} \wedge \cdots \wedge dx^{\mu_n}$$

The coefficients $L^{\mu_1\cdots\mu_k|\alpha_1\cdots\alpha_k}$ have to obey certain *structure equations* (see [1] for more details) but in principle admit the following explicit local representation in terms of ordinary Lagrange function:

$$L^{\mu_1,\ldots\mu_k|\alpha_1,\ldots\alpha_k} = \frac{1}{k!} \sum_{\varsigma\in P_k} (-1)^{|\varsigma|} \frac{\partial^k \boldsymbol{\mathcal{L}}}{\partial\chi_{\alpha_1\mu_{\varsigma(1)}} \ldots \partial\chi_{\alpha_k\mu_{\varsigma(k)}}}.$$

$P_k$ is the permutation group of numbers $1, \ldots, k$ and $|\varsigma|$ is the signature of $\varsigma$. $\boldsymbol{\mathcal{L}}$ is the ordinary Lagrange function.

For the vacuum electromagnetic field it is convenient to choose the Lagrange function in the form

$$\boldsymbol{\mathcal{L}} = 2\chi_{\alpha\mu}\chi_{[\beta\nu]}\sqrt{-g}g^{\alpha\beta}g^{\mu\nu}$$

(if $\chi_{\alpha\mu}$ are interpreted as $\frac{\partial\psi_\alpha}{\partial x^\mu}$ it will be reduced to the usual representation) and we have respectively

$$\begin{aligned}
\theta =& \epsilon_{\mu_1\mu_2\mu_3\mu_4} \sum_{k=0}^{4} \frac{1}{k!} C_4^k L^{\mu_1\cdots\mu_k|\alpha_1\cdots\alpha_k} \delta\psi_{\alpha_1} \wedge \cdots \wedge \delta\psi_{\alpha_k} \wedge dx^{\mu_{k+1}} \wedge \cdots \wedge dx^{\mu_4} \\
=& \epsilon_{\mu_1\mu_2\mu_3\mu_4} (1 \cdot L\, dx^{\mu_1} \wedge dx^{\mu_2} \wedge dx^{\mu_3} \wedge dx^{\mu_4} \\
& \qquad \frac{4}{1!} \cdot L^{\mu_1|\alpha_1}\, \delta\psi_{\alpha_1} \wedge dx^{\mu_2} \wedge dx^{\mu_3} \wedge dx^{\mu_4} \\
& \qquad \frac{6}{2!} \cdot L^{\mu_1\mu_2|\alpha_1\alpha_2}\, \delta\psi_{\alpha_1} \wedge \delta\psi_{\alpha_2} \wedge dx^{\mu_3} \wedge dx^{\mu_4}) \\
=& \boldsymbol{\mathcal{L}}\,\eta + 4L^{\mu_1|\alpha_1}\, \delta\psi_{\alpha_1} \wedge \eta_{\mu_1} + 3L^{\mu_1\mu_2|\alpha_1\alpha_2}\, \delta\psi_{\alpha_1} \wedge \delta\psi_{\alpha_2} \wedge \eta_{\mu_1\mu_2}
\end{aligned}$$

where we have denoted for convenience

$$\begin{aligned}
\eta_{\mu_1\mu_2} &= \epsilon_{\mu_1\mu_2\mu_3\mu_4} dx^{\mu_3} \wedge dx^{\mu_4}, \\
\eta_{\mu_1} &= \epsilon_{\mu_1\mu_2\mu_3\mu_4} dx^{\mu_2} \wedge dx^{\mu_3} \wedge dx^{\mu_4} = dx^{\mu_2} \wedge \eta_{\mu_1\mu_2}, \\
\eta &= \epsilon_{\mu_1\mu_2\mu_3\mu_4} dx^{\mu_1} \wedge dx^{\mu_2} \wedge dx^{\mu_3} \wedge dx^{\mu_4} = dx^{\mu_1} \wedge \eta_{\mu_1}.
\end{aligned}$$

Other $L$'s vanish. One has further

$$L = \mathcal{L},$$

$$\begin{aligned} L^{\mu_1|\alpha_1} &= \frac{\partial \mathcal{L}}{\partial \chi_{\alpha_1\mu_1}} = \frac{\partial}{\partial \chi_{\alpha_1\mu_1}} \; 2\chi_{\alpha\mu}\chi_{\beta\nu}\sqrt{-g}g^{\alpha[\beta}g^{\nu]\mu} \\ &= 2(\chi_{\beta\nu}\sqrt{-g}g^{\alpha_1[\beta}g^{\nu]\mu_1} + \chi_{\alpha\mu}\sqrt{-g}g^{\alpha[\alpha_1}g^{\mu_1]\mu}) \\ &= 4\chi_{[\beta\nu]}\sqrt{-g}g^{\alpha_1\beta}g^{\nu\mu_1}; \end{aligned}$$

$$\begin{aligned} L^{\mu_1\mu_2|\alpha_1\alpha_2} &= \tfrac{1}{2}\left(\frac{\partial^2 \mathcal{L}}{\partial \chi_{\alpha_1\mu_1}\chi_{\alpha_2\mu_2}} - \frac{\partial^2 \mathcal{L}}{\partial \chi_{\alpha_1\mu_2}\chi_{\alpha_2\mu_1}}\right) \\ &= \tfrac{1}{2}\left(\frac{\partial}{\partial \chi_{\alpha_1\mu_1}} \; L^{\mu_2|\alpha_2} - \frac{\partial}{\partial \chi_{\alpha_1\mu_2}} \; L^{\mu_1|\alpha_2}\right). \end{aligned}$$

Meanwhile

$$\frac{\partial}{\partial \chi_{\alpha_1\mu_1}} \; L^{\mu_2|\alpha_2} = \frac{\partial}{\partial \chi_{\alpha_1\mu_1}} \; 4\chi_{\beta\nu}\sqrt{-g}g^{\alpha_2[\beta}g^{\nu]\mu_2} = 4\sqrt{-g}g^{\alpha_2[\alpha_1}g^{\mu_1]\mu_2}$$

and therefore

$$\begin{aligned} L^{\mu_1\mu_2|\alpha_1\alpha_2} &= 2\left(\sqrt{-g}g^{\alpha_2[\alpha_1}g^{\mu_1]\mu_2} - \sqrt{-g}g^{\alpha_2[\alpha_1}g^{\mu_2]\mu_1}\right) \\ &= \underline{\sqrt{-g}g^{\alpha_2\alpha_1}g^{\mu_1\mu_2}} - \sqrt{-g}g^{\alpha_2\mu_1}g^{\alpha_1\mu_2} - \underline{\sqrt{-g}g^{\alpha_2\alpha_1}g^{\mu_2\mu_1}} + \sqrt{-g}g^{\alpha_2\mu_2}g^{\alpha_1\mu_1} \\ &= \sqrt{-g}g^{\alpha_1\mu_1}g^{\alpha_2\mu_2} - \sqrt{-g}g^{\alpha_1\mu_2}g^{\mu_1\alpha_2} \\ &= 2\sqrt{-g}g^{\alpha_1[\mu_1}g^{\mu_2]\alpha_2}. \end{aligned}$$

Thus

$$\begin{aligned} \theta = \; & 2\sqrt{-g}g^{\alpha[\beta}g^{\nu]\mu}\,\chi_{\alpha\mu}\chi_{\beta\nu}\,\eta \\ & + 16\sqrt{-g}g^{\alpha[\beta}g^{\nu]\mu}\chi_{\beta\nu}\,\delta\psi_\alpha \wedge \eta_\mu \\ & + 6\sqrt{-g}g^{\alpha_1\mu_1}g^{\mu_2\alpha_2}\,\delta\psi_{\alpha_1} \wedge \delta\psi_{\alpha_2} \wedge \eta_{\mu_1\mu_2} \end{aligned}$$

It is essential to mention that this generalized Cartan form contains, besides antisymmetric part of $\chi_{\alpha\mu}$, its symmetric part as well.

The Euler-Lagrange equations in frames of the formalism under consideration take the form

$$\Phi^{1^*}(X^1 \lrcorner\, \sigma) = 0$$

which equation has to be observed for *arbitrary* vector field $X$ on the evolution space. Here $X^1$ denotes its lift to the jet-1 bundle, $\Phi^{1^*}$ is the pull-back of the immersion $\Phi^1$ of the space-time into the jet-1 bundle which is described in the coordinates $\{x^\mu, \psi_\alpha, \chi_{\alpha\mu}\}$ by the equations

$$\begin{aligned} x^\mu &= x^\mu, \\ \psi_\alpha &= \Psi_\alpha(x), \\ \chi_{\alpha\mu} &= \frac{\partial \Psi_\alpha}{\partial x^\mu}(x) \end{aligned}$$

for some functions $\Psi_\alpha(x)$. The latter determines the solutions of the Euler-Lagrange equations if and only if after such a 'substitution' the $N$-form ($N$ is the dimension of space-time) $X^1 \rfloor \sigma$ vanishes identically for every $X$ (the pull-back of the differential form is locally nothing else but the result of substitutions of the representations of coordinates of the space 'to where' in terms of coordinates 'from where').

Now we shall calculate the lift of the vector field $X$ to the jet-1 bundle following the work [6]. It can be done as follows.

The vector $X$ can be represented in the form

$$X = \xi^\mu(x) \frac{\partial}{\partial x^\mu} + \zeta_\alpha(x, \psi) \frac{\partial}{\partial \psi_\alpha}$$

($\xi$ and $\zeta$ are arbitrary functions). This vector field is not of the most general form since $\xi$'s do not depend on $\psi_\alpha$. But this is necessary restriction because otherwise infinitesimal transformation of evolution space produced by $X$ (see below) would introduce into the functions on space-time ($g_{\alpha\beta}$ in our case) *explicit* dependence on field variables $\psi_\alpha$.

The vector field $X$ on the evolution space is associated with the infinitesimal motions of the space in accordance with equations

$$\begin{cases} x^\mu \to x^\mu + \epsilon \xi^\mu(x) + o(\epsilon) & = \tilde{x}^\mu + o(\epsilon), \\ \psi_\alpha \to \psi_\alpha + \epsilon \zeta_\alpha(x, \psi) + o(\epsilon) & = \tilde{\psi}_\alpha + o(\epsilon). \end{cases}$$

If $\chi_{\alpha\mu}$ were interpreted by means of equation

$$d\psi_\alpha = \chi_{\alpha\mu} dx^\mu$$

then the 'lifted' *transformed* $\chi_{\alpha\mu}$ (*i.e.* $\tilde{\chi}_{\alpha\mu}$) were defined by means of a similar equation

$$d\tilde{\psi}_\alpha = \tilde{\chi}_{\alpha\mu} d\tilde{x}^\mu.$$

Using the explicit form of the expressions for $\tilde{x}^\mu, \tilde{\psi}_\alpha$ one can find the following transformation law of $\chi_{\alpha\mu}$:

$$\tilde{\chi}_{\alpha\mu} = \chi_{\alpha\mu} + \epsilon \left( \frac{\partial \zeta_\alpha}{\partial x^\mu} + \frac{\partial \zeta_\alpha}{\partial \psi_\beta} \chi_{\beta\mu} - \frac{\partial \xi^\nu}{\partial x^\mu} \chi_{\alpha\nu} \right) + o(\epsilon)$$

The complete transformation expressed above in terms of jet bundle coordinates $\{x^\mu, \psi_\alpha, \chi_{\alpha\mu}\}$ corresponds to the vector

$$X^1 = \xi^\mu(x) \frac{\partial}{\partial x^\mu} + \zeta_\alpha(x, \psi) \frac{\partial}{\partial \psi_\alpha} + \left( \frac{\partial \zeta_\alpha}{\partial x^\mu} + \frac{\partial \zeta_\alpha}{\partial \psi_\beta} \chi_{\beta\mu} - \frac{\partial \xi^\nu}{\partial x^\mu} \chi_{\alpha\nu} \right) \frac{\partial}{\partial \chi_{\alpha\mu}}$$

which is just the lift $X^1$ of the vector $X$ to the jet-1 bundle. It turns out to be convenient to represent it as a sum of two 'independent' vector fields

$$X^1 = X^1_\xi + X^1_\zeta$$

where

$$\begin{aligned} X^1_\xi &= \xi^\mu(x)\frac{\partial}{\partial x^\mu} - \frac{\partial \xi^\nu}{\partial x^\mu}\chi_{\alpha\nu}\frac{\partial}{\partial \chi_{\alpha\mu}} \\ X^1_\zeta &= \zeta_\alpha(x,\psi)\frac{\partial}{\partial \psi_\alpha} + \left(\frac{\partial \zeta_\alpha}{\partial x^\mu} + \frac{\partial \zeta_\alpha}{\partial \psi_\beta}\chi_{\beta\mu}\right)\frac{\partial}{\partial \chi_{\alpha\mu}} \end{aligned}$$

It is clear that Euler-Lagrange equations must be obeyed with ether $X^1 = X^1_\xi$ or $X^1 = X^1_\zeta$ choices independently.

Now we need calculate $\sigma = d\theta$ using the explicit representation of $\theta$ given above. The following obvious equations are useful:

$$d\delta\psi_\alpha = -d\chi_{\alpha\mu} \wedge dx^\mu, \qquad d\eta = d\eta_\mu = d\eta_{\mu_1\mu_2} = 0.$$

One has

$$\begin{aligned} d\theta = \ & 4\sqrt{-g}g^{\alpha[\beta}g^{\nu]\mu}\,\chi_{\beta\nu}\,d\chi_{\alpha\mu}\wedge\eta \\ & - 16\{\sqrt{-g}g^{\alpha_1[\beta}g^{\nu]\mu_1}\}_{,\mu_0}\,\chi_{\beta\nu}\,\delta\psi_{\alpha_1}\wedge dx^{\mu_0}\wedge\eta_{\mu_1} \\ & + 16\sqrt{-g}g^{\alpha_1[\beta}g^{\nu]\mu_1}\,d\chi_{\beta\nu}\wedge\delta\psi_{\alpha_1}\wedge\eta_{\mu_1} \\ & - 16\sqrt{-g}g^{\alpha_1[\beta}g^{\nu]\mu_1}\,\chi_{\beta\nu}\,d\chi_{\alpha_1\mu_0}\wedge dx^{\mu_0}\wedge\eta_{\mu_1} \\ & + 6\{\sqrt{-g}g^{\alpha_1\mu_1}g^{\mu_2\alpha_2}\}_{,\mu_0}\,\delta\psi_{\alpha_1}\wedge\delta\psi_{\alpha_2}\wedge dx^{\mu_0}\wedge\eta_{\mu_1\mu_2} \\ & + 12\sqrt{-g}g^{\alpha_1\mu_1}g^{\mu_2\alpha_2}\,d\chi_{\alpha_1\mu_0}\wedge\delta\psi_{\alpha_2}\wedge dx^{\mu_0}\wedge\eta_{\mu_1\mu_2}. \end{aligned}$$

Some simplifications can be achieved by means of equations

$$\begin{aligned} dx^{\mu_0}\wedge\eta_{\mu_1} &= \tfrac{1}{4}\delta^{\mu_0}_{\mu_1}\eta, \\ dx^{\mu_0}\wedge\eta_{\mu_1\mu_2} &= \tfrac{1}{3}(\delta^{\mu_0}_{\mu_1}\eta_{\mu_2} - \delta^{\mu_0}_{\mu_2}\eta_{\mu_1}). \end{aligned}$$

They yield

$$\begin{aligned} d\theta = \ & 4\,\underline{\sqrt{-g}g^{\alpha[\beta}g^{\nu]\mu}\,\chi_{\beta\nu}\,d\chi_{\alpha\mu}\wedge\eta} \\ & - 4\{\sqrt{-g}g^{\alpha_1[\beta}g^{\nu]\mu}\}_{,\mu}\,\chi_{\beta\nu}\,\delta\psi_{\alpha_1}\wedge\eta \\ & + 16\sqrt{-g}g^{\alpha_1[\beta}g^{\nu]\mu_1}\,d\chi_{\beta\nu}\wedge\delta\psi_{\alpha_1}\wedge\eta_{\mu_1} \end{aligned}$$

$$
\begin{aligned}
&- 4\, \underline{\sqrt{-g} g^{\alpha[\beta} g^{\nu]\mu}\, \chi_{\beta\nu}\, d\chi_{\alpha\mu} \wedge \eta} \\
&+ 4\{\sqrt{-g} g^{\alpha_1[\mu} g^{\mu_2]\alpha_2}\}_{,\mu}\; \delta\psi_{\alpha_1} \wedge \delta\psi_{\alpha_2} \wedge \eta_{\mu_2} \\
&+ 8\, \sqrt{-g} g^{\alpha_1[\mu} g^{\mu_2]\alpha_2}\, d\chi_{\alpha_1\mu} \wedge \delta\psi_{\alpha_2} \wedge \eta_{\mu_2}
\end{aligned}
$$

$$
\begin{aligned}
= &- 4\{\sqrt{-g} g^{\alpha[\beta} g^{\nu]\mu}\}_{,\mu}\, \chi_{\beta\nu}\, \delta\psi_{\alpha} \wedge \eta \\
&+ 16\, \sqrt{-g} g^{\alpha[\beta} g^{\nu]\mu}\, d\chi_{\beta\nu} \wedge \delta\psi_{\alpha} \wedge \eta_{\mu} \\
&+ 4\{\sqrt{-g} g^{\beta[\nu} g^{\mu]\alpha}\}_{,\nu}\; \delta\psi_{\beta} \wedge \delta\psi_{\alpha} \wedge \eta_{\mu} \\
&+ 8\, \sqrt{-g} g^{\beta[\nu} g^{\mu]\alpha}\, d\chi_{\beta\nu} \wedge \delta\psi_{\alpha} \wedge \eta_{\mu}.
\end{aligned}
$$

This is an explicit representation of Lagrange-Souriau form for electromagnetic field. As well as generalized Cartan form it involves symmetric part of $\chi_{\alpha\mu}$ because entering the last term $d\chi_{(\beta\nu)}$ certainly cannot be cancelled out.

Now we can calculate the contractions $Z^1_\xi \rfloor \sigma$ and $Z^1_\zeta \rfloor \sigma$. It must be emphasized however that these contractions produce the Euler-Lagrange equations *when they are restricted to the image of the space-time* under the immersion $\Phi$. The latter however possesses the property to annihilate $\delta\psi_\alpha$: ${\Phi^1}^*(\delta\psi_\alpha) = 0$. Thus, in order to construct the Euler-Lagrange equations it is sufficient to keep only those terms of the contractions that do *not* contain the factor $\delta\psi_\alpha$. But all the terms of $d\theta$ representation do contain it at least once. Hence, the vectors $Z^1_\xi$ or $Z^1_\zeta$ have to be contracted just with such a factor and the term may not contain other factors $\delta\psi_\alpha$ if it contributes the resulting equations.

One has obviously

$$
\begin{aligned}
Z^1_\xi \rfloor d\chi_{\alpha\mu} &= -\frac{\partial\xi^\nu}{\partial x^\mu}\chi_{\alpha\nu} \\
Z^1_\xi \rfloor \delta\psi_\alpha &= -\xi^\mu \chi_{\alpha\mu} \\
Z^1_\xi \rfloor \eta_\mu &= 3\xi^\nu \eta_{\mu\nu}.
\end{aligned}
$$

Similarly

$$
\begin{aligned}
Z^1_\zeta \rfloor d\chi_{\alpha\mu} &= \frac{\partial\zeta_\alpha}{\partial x^\mu} + \frac{\partial\zeta_\alpha}{\partial\psi_\beta}\chi_{\beta\mu} \\
Z^1_\zeta \rfloor \delta\psi_\alpha &= \zeta_\alpha(x,\psi) \\
Z^1_\zeta \rfloor \eta_\mu &= 0.
\end{aligned}
$$

It follows also from these equations and the above remark that in our case

$$
{\Phi^1}^*(Z^1_\zeta \rfloor \sigma) \equiv 0 \qquad \Longrightarrow \qquad {\Phi^1}^*(Z^1_\xi \rfloor \sigma) \equiv 0.
$$

and therefore it is sufficient to calculate $Z^1_\zeta \rfloor \sigma \quad mod\ \delta\psi_\alpha$.

The contraction with Lagrange-Souriau form $\sigma = d\theta$ will be the following:

$$\begin{aligned} Z^1_\zeta \rfloor \sigma \overset{[\mathrm{mod}\ \delta\psi_\alpha]}{=} & - \zeta_\alpha\, 4\{\sqrt{-g} g^{\alpha[\beta} g^{\nu]\mu}\}_{,\mu}\, \chi_{\beta\nu}\, \eta \\ & - \zeta_\alpha\, 16 \sqrt{-g} g^{\alpha[\beta} g^{\nu]\mu}\, d\chi_{\beta\nu} \wedge \eta_\mu \\ & - \zeta_\alpha\, 8 \sqrt{-g} g^{\beta[\nu} g^{\mu]\alpha}\, d\chi_{\beta\nu} \wedge \eta_\mu . \end{aligned}$$

It remains to substitute $\chi_{\beta\nu} = \partial_\nu \Psi_\beta(x)$ to obtain the following version

$$\begin{aligned} & \{\sqrt{-g} g^{\alpha[\beta} g^{\nu]\mu}\}_{,\mu} \partial_\nu \Psi_\beta\, \eta \\ & +4 \sqrt{-g} g^{\alpha[\beta} g^{\nu]\mu}\, d\partial_\nu \Psi_\beta \wedge \eta_\mu \\ & +2 \sqrt{-g} g^{\beta[\nu} g^{\mu]\alpha}\, d\partial_\nu \Psi_\beta \wedge \eta_\mu = 0 \end{aligned}$$

of the vacuum Maxwell equation.

Is it a true one? It is not so obvious at the moment.

Let us note however that the last term of the above equation vanishes identically because

$$d\partial_\nu \Psi_\beta \wedge \eta_\mu = \partial_{\rho\nu} \Psi_\beta\, dx^\rho \wedge \eta_\mu = \tfrac{1}{4} \partial_{\mu\nu} \Psi_\beta\, \eta$$

is symmetric in $\mu, \nu$ and the factor $g^{\beta[\nu} g^{\mu]\alpha}$ is antisymmetric. Then the implied equation can be expressed in term of antisymmetric E-M stress tensor

$$F_{\beta\nu} = \partial_{[\nu} \Psi_{\beta]}$$

alone simultaneously removing antisymmetrizations from the metric tensor aggregates:

$$\begin{aligned} & \{\sqrt{-g} g^{\alpha\beta} g^{\nu\mu}\}_{,\mu}\, F_{\beta\nu}\, \eta + 4 \sqrt{-g} g^{\alpha\beta} g^{\nu\mu}\, dF_{\beta\nu} \wedge \eta_\mu = 0 \qquad \Rightarrow \\ & \{\sqrt{-g} g^{\alpha\beta} g^{\nu\mu}\}_{,\mu}\, F_{\beta\nu} + \sqrt{-g} g^{\alpha\beta} g^{\nu\mu}\, \partial_\mu F_{\beta\nu} = \\ & g_{\alpha\gamma} \{\sqrt{-g} g^{\alpha\beta} g^{\nu\mu}\}_{,\mu}\, F_{\beta\nu} + \sqrt{-g} g^{\nu\mu}\, \partial_\mu F_{\gamma\nu} = \\ & g_{\alpha\gamma} \{\sqrt{-g} g^{\alpha\beta} g^{\nu\mu} F_{\beta\nu}\}_{,\mu} \\ & \quad - \underline{\sqrt{-g} g_{\alpha\gamma} g^{\alpha\beta} g^{\nu\mu} \partial_\mu F_{\beta\nu}} + \underline{\sqrt{-g} g^{\nu\mu}\, \partial_\mu F_{\beta\nu}} = \\ & g_{\alpha\gamma} \{\sqrt{-g} F^{\alpha\mu}\}_{,\mu} = 0 \end{aligned}$$

where we have defined as usually $F^{\alpha\mu} \equiv g^{\alpha\beta} g^{\nu\mu} F_{\beta\nu}$. This is just the vacuum Maxwell equations

$$\nabla_\mu F^{\alpha\mu} = 0$$

that follows from the wellknown definition of covariant derivative:

$$\begin{aligned} \partial_\mu F^{\alpha\nu} &= \nabla_\mu F^{\alpha\nu} - \Gamma^\alpha_{\mu\rho} F^{\rho\nu} - \Gamma^\nu_{\mu\rho} F^{\rho\alpha} \qquad \Rightarrow \\ \partial_\mu F^{\alpha\mu} &= \nabla_\mu F^{\alpha\mu} - \Gamma^\mu_{\mu\rho} F^{\rho\alpha}, \\ \Gamma^\mu_{\mu\rho} &= \frac{1}{\sqrt{-g}} \{\sqrt{-g}\}_{,\rho}. \end{aligned}$$

We have obtained Maxwell equations in 'apparently covariant' (with respect to coordinate transformations) form. The general definition of $\theta$ can be seen as 'apparently covariant' as well. Meanwhile the above expression of the Lagrange-Souriau form obtained by straightforward calculation does not reveal such a feature. It makes sense to obtain an 'apparently covariant' representation of $\sigma$ also.

Let us transform $\sigma$

$$
\begin{aligned}
\tfrac{1}{4}\sigma = &- \left\{\sqrt{-g}g^{\alpha[\beta}g^{\nu]\mu}\right\}_{,\mu}\chi_{\beta\nu}\,\delta\psi_\alpha\wedge\eta \\
&+ \left\{\sqrt{-g}g^{\beta[\nu}g^{\mu]\alpha}\right\}_{,\nu}\,\delta\psi_\beta\wedge\delta\psi_\alpha\wedge\eta_\mu \\
&+ \left\{4\sqrt{-g}g^{\alpha[\beta}g^{\nu]\mu} + 2\sqrt{-g}g^{\beta[\nu}g^{\mu]\alpha}\right\}d\chi_{\beta\nu}\wedge\delta\psi_\alpha\wedge\eta_\mu
\end{aligned}
$$

eliminating derivatives of metric tensor by means of formulae

$$
\{\sqrt{-g}\}_{,\mu} = \sqrt{-g}\,\Gamma^\rho_{\rho\mu}, \qquad g^{\alpha\beta}{}_{,\mu} = -\Gamma^\alpha_{\mu\rho}\,g^{\rho\beta} - \Gamma^\beta_{\mu\rho}\,g^{\rho\alpha}.
$$

One has

$$
\begin{aligned}
\left\{\sqrt{-g}g^{\alpha[\beta}g^{\nu]\mu}\right\}_{,\mu} &= \sqrt{-g}\Big( + \underline{\underline{\Gamma^\rho_{\rho\mu}g^{\alpha[\beta}g^{\nu]\mu}}} \\
&\qquad - \underline{\Gamma^\alpha_{\mu\rho}g^{\rho[\beta}g^{\nu]\mu}} - g^{\alpha\rho}\Gamma^{[\beta}_{\mu\rho}g^{\nu]\mu} \\
&\qquad - g^{\alpha[\beta}\Gamma^{\nu]}_{\mu\rho}g^{\mu\rho} - \underline{\underline{\Gamma^\mu_{\mu\rho}g^{\alpha[\beta}g^{\nu]\rho}}}\Big) \\
&= \sqrt{-g}\Big( - g^{\alpha\rho}g^{\mu[\nu}\Gamma^{\beta]}_{\mu\rho} - g^{\rho\mu}g^{\alpha[\beta}\Gamma^{\nu]}_{\mu\rho}\Big) \\
&= -\sqrt{-g}g^{\rho[\alpha}g^{\mu][\nu}\Gamma^{\beta]}_{\mu\rho},
\end{aligned}
$$

similarly

$$
\begin{aligned}
\left\{\sqrt{-g}g^{\beta[\nu}g^{\mu]\alpha}\right\}_{,\nu} &= \left\{\sqrt{-g}g^{\nu[\beta}g^{\alpha]\mu}\right\}_{,\nu} \\
&= \sqrt{-g}\Big( \underline{\underline{\Gamma^\rho_{\rho\nu}g^{\nu[\beta}g^{\alpha]\mu}}} \\
&\qquad - g^{\mu[\alpha}\Gamma^{\beta]}_{\nu\rho}g^{\nu\rho} - \underline{\underline{\Gamma^\nu_{\nu\rho}g^{\rho[\beta}g^{\alpha]\mu}}} \\
&\qquad - \underline{\Gamma^\mu_{\rho\nu}g^{\nu[\beta}g^{\alpha]\rho}} - g^{\rho\mu}g^{\nu[\beta}\Gamma^{\alpha]}_{\rho\nu}\Big) \\
&= -2\sqrt{-g}g^{\rho[\nu}g^{\mu][\alpha}\Gamma^{\beta]}_{\nu\rho}
\end{aligned}
$$

and therefore

$$
\begin{aligned}
\frac{\sigma}{8\sqrt{-g}} = \;& g^{\rho[\alpha}g^{\mu][\nu}\Gamma^{\beta]}_{\mu\rho}\chi_{\beta\nu}\,\delta\psi_\alpha\wedge\eta \\
&- g^{\rho[\nu}g^{\mu]\alpha}\Gamma^\beta_{\nu\rho}\delta\psi_\beta\wedge\delta\psi_\alpha\wedge\eta_\mu \\
&+ \left\{2\,g^{\alpha[\beta}g^{\nu]\mu} + g^{\beta[\nu}g^{\mu]\alpha}\right\}d\chi_{\beta\nu}\wedge\delta\psi_\alpha\wedge\eta_\mu.
\end{aligned}
$$

Next we introduce "formally covariant" objects

$$X_{\alpha\mu} = \chi_{\alpha\mu} - \Gamma^{\beta}_{\alpha\mu}\psi_\beta$$

and their "formally covariant differential"

$$D\,X_{\alpha\mu} = dX_{\alpha\mu} - X_{\beta\mu}\,\Gamma^{\beta}_{\alpha\rho}\,dx^\rho - X_{\alpha\nu}\,\Gamma^{\nu}_{\mu\rho}\,dx^\rho.$$

One has therefore

$$d\chi_{\alpha\mu} = dX_{\alpha\mu} + \Gamma^{\beta}_{\alpha\mu}\,d\psi_\beta + \psi_\beta\,d\Gamma^{\beta}_{\alpha\mu},$$

where it is necessary to substitute for

$$d\psi_\beta = \delta\psi_\beta + \chi_{\beta\rho}\,dx^\rho = \delta\psi_\beta + X_{\beta\rho}\,dx^\rho + \psi_\gamma\Gamma^{\gamma}_{\beta\rho}\,dx^\rho.$$

It is worthwhile to mention that

$$\chi_{[\alpha\mu]} = X_{[\alpha\mu]}, \qquad d\chi_{[\alpha\mu]} = dX_{[\alpha\mu]} = DX_{[\alpha\mu]} + X_{[\alpha\nu]}\Gamma^{\nu}_{\mu\rho}dx^\rho + X_{[\beta\mu]}\Gamma^{\beta}_{\alpha\rho}dx^\rho$$

and one has for

$$\begin{aligned}
2\,g^{\alpha[\beta}g^{\nu]\mu}\,d\chi_{\beta\nu}\wedge\delta\psi_\alpha\wedge\eta_\mu =& 2\,g^{\alpha\beta}g^{\nu\mu}\Big(DX_{[\beta\nu]} + X_{[\beta\sigma]}\Gamma^{\sigma}_{\nu\rho}dx^\rho + X_{[\gamma\nu]}\Gamma^{\gamma}_{\beta\rho}dx^\rho\Big)\wedge\delta\psi_\alpha\wedge\eta_\mu\\
=& 2\,g^{\alpha\beta}g^{\nu\mu}DX_{[\beta\nu]}\wedge\delta\psi_\alpha\wedge\eta_\mu - \tfrac{1}{2}g^{\alpha\beta}g^{\nu\mu}X_{[\beta\sigma]}\Gamma^{\sigma}_{\nu\mu}\delta\psi_\alpha\wedge\eta\\
&- \tfrac{1}{2}g^{\alpha\beta}g^{\nu\mu}X_{[\gamma\nu]}\Gamma^{\gamma}_{\beta\mu}\delta\psi_\alpha\wedge\eta\\
=& 2\,g^{\alpha\beta}g^{\nu\mu}DX_{[\beta\nu]}\wedge\delta\psi_\alpha\wedge\eta_\mu - \tfrac{1}{2}g^{\alpha\beta}g^{\nu\mu}X_{[\beta\sigma]}\Gamma^{\sigma}_{\nu\mu}\delta\psi_\alpha\wedge\eta\\
&- \tfrac{1}{2}g^{\alpha\nu}g^{\beta\mu}X_{[\sigma\beta]}\Gamma^{\sigma}_{\nu\mu}\delta\psi_\alpha\wedge\eta\\
=& 2\,g^{\alpha\beta}g^{\nu\mu}DX_{[\beta\nu]}\wedge\delta\psi_\alpha\wedge\eta_\mu - g^{\alpha[\beta}g^{\nu]\mu}X_{[\beta\sigma]}\Gamma^{\sigma}_{\nu\mu}\delta\psi_\alpha\wedge\eta.
\end{aligned}$$

Meanwhile without antisymmetrization one has

$$d\chi_{\beta\nu} = D\,X_{\beta\nu} + \Gamma^{\gamma}_{\beta\nu}\,\delta\psi_\gamma + \{\psi_\gamma\,\partial_\rho\Gamma^{\gamma}_{\beta\nu} + X_{\gamma\rho}\,\Gamma^{\gamma}_{\beta\nu} + \psi_\delta\,\Gamma^{\delta}_{\gamma\rho}\Gamma^{\gamma}_{\beta\nu} + X_{\gamma\nu}\,\Gamma^{\gamma}_{\beta\rho} + X_{\beta\sigma}\,\Gamma^{\sigma}_{\nu\rho}\}\,dx^\rho$$

and therefore

$$\begin{aligned}
d\chi_{\beta\nu}\wedge\delta\psi_\alpha\wedge\eta_\mu =& \{D\,X_{\beta\nu} + \Gamma^{\gamma}_{\beta\nu}\,\delta\psi_\gamma\}\wedge\delta\psi_\alpha\wedge\eta_\mu\\
&- \{\psi_\gamma\,\partial_\rho\Gamma^{\gamma}_{\beta\nu} + X_{\gamma\rho}\,\Gamma^{\gamma}_{\beta\nu} + \psi_\delta\,\Gamma^{\delta}_{\gamma\rho}\Gamma^{\gamma}_{\beta\nu} + X_{\gamma\nu}\,\Gamma^{\gamma}_{\beta\rho} + X_{\beta\sigma}\,\Gamma^{\sigma}_{\nu\rho}\}\times\\
&\qquad\qquad\qquad\qquad\qquad\qquad\qquad\qquad\delta\psi_\alpha\wedge dx^\rho\wedge\eta_\mu\\
=& \{D\,X_{\beta\nu} + \Gamma^{\gamma}_{\beta\nu}\,\delta\psi_\gamma\}\wedge\delta\psi_\alpha\wedge\eta_\mu\\
&- \tfrac{1}{4}\{\psi_\gamma\,\partial_\mu\Gamma^{\gamma}_{\beta\nu} + \psi_\gamma\,\Gamma^{\gamma}_{\delta\mu}\Gamma^{\delta}_{\beta\nu} + 2\,X_{\gamma(\mu}\,\Gamma^{\gamma}_{\nu)\beta} + X_{\beta\sigma}\,\Gamma^{\sigma}_{\nu\mu}\}\delta\psi_\alpha\wedge\eta,
\end{aligned}$$

$$\begin{aligned}
g^{\beta[\nu}g^{\mu]\alpha}\,d\chi_{\beta\nu}\wedge\delta\psi_\alpha\wedge\eta_\mu =& g^{\beta[\nu}g^{\mu]\alpha}\,D\,X_{\beta\nu}\wedge\delta\psi_\alpha\wedge\eta_\mu + g^{\beta[\nu}g^{\mu]\alpha}\Gamma^{\gamma}_{\beta\nu}\,\delta\psi_\gamma\wedge\delta\psi_\alpha\wedge\eta_\mu\\
&- \tfrac{1}{4}g^{\beta\nu}g^{\mu\alpha}\{\psi_\gamma\,\partial_{[\mu}\Gamma^{\gamma}_{\nu]\beta} + \psi_\gamma\,\Gamma^{\gamma}_{\delta[\mu}\Gamma^{\delta}_{\nu]\beta}\}\delta\psi_\alpha\wedge\eta.
\end{aligned}$$

The latter expression contains precisely the Riemann tensor

$$R^{\gamma}_{\beta\nu\mu} = 2\,\partial_{[\mu}\Gamma^{\gamma}_{\nu]\beta} + 2\,\Gamma^{\gamma}_{\delta[\mu}\Gamma^{\delta}_{\nu]\beta}$$

and, after contraction with metric, the Ricci tensor $R^{\gamma\alpha} = R^{\gamma}{}_{\nu}{}^{\alpha\nu}$ enter it as follows

$$\begin{aligned} g^{\beta[\nu}g^{\mu]\alpha}\, d\chi_{\beta\nu} \wedge \delta\psi_\alpha \wedge \eta_\mu =& g^{\beta[\nu}g^{\mu]\alpha}\, D\,X_{\beta\nu} \wedge \delta\psi_\alpha \wedge \eta_\mu + g^{\beta[\nu}g^{\mu]\alpha}\Gamma^{\gamma}_{\beta\nu}\,\delta\psi_\gamma \wedge \delta\psi_\alpha \wedge \eta_\mu \\ &+ \tfrac{1}{8}\psi_\gamma R^{\gamma\alpha}\delta\psi_\alpha \wedge \eta. \end{aligned}$$

As a result one has the following expression of the the Lagrange-Souriau form

$$\begin{aligned} \frac{\sigma}{8\sqrt{-g}} =& \underline{g^{\rho[\alpha}g^{\mu]\nu}\Gamma^{\beta}_{\mu\rho}X_{[\beta\nu]}\,\delta\psi_\alpha \wedge \eta} \\ &- \underline{\underline{g^{\rho[\nu}g^{\mu]\alpha}\Gamma^{\beta}_{\nu\rho}\delta\psi_\beta \wedge \delta\psi_\alpha \wedge \eta_\mu}} \\ &+ 2\,g^{\alpha\beta}g^{\nu\mu}\,D\,X_{[\beta\nu]} \wedge \delta\psi_\alpha \wedge \eta_\mu - \underline{g^{\alpha[\beta}g^{\nu]\mu}\,\Gamma^{\sigma}_{\mu\nu}\,X_{\beta\sigma}\delta\psi_\alpha \wedge \eta} \\ &+ g^{\beta[\nu}g^{\mu]\alpha}\,D\,X_{\beta\nu} \wedge \delta\psi_\alpha \wedge \eta_\mu + \underline{\underline{g^{\beta[\nu}g^{\mu]\alpha}\Gamma^{\gamma}_{\beta\nu}\,\delta\psi_\gamma \wedge \delta\psi_\alpha \wedge \eta_\mu}} \\ &+ \tfrac{1}{8}\psi_\gamma R^{\gamma\alpha}\delta\psi_\alpha \wedge \eta. \end{aligned}$$

Underlined terms are mutually cancelled and

$$\begin{aligned} \frac{\sigma}{8\sqrt{-g}} =& \tfrac{1}{8}\psi_\gamma R^{\gamma\alpha}\delta\psi_\alpha \wedge \eta \\ &+ 2\,g^{\alpha\beta}g^{\nu\mu}\,D\,X_{[\beta\nu]} \wedge \delta\psi_\alpha \wedge \eta_\mu + g^{\beta[\nu}g^{\mu]\alpha}\,D\,X_{\beta\nu} \wedge \delta\psi_\alpha \wedge \eta_\mu \end{aligned}$$

Let us consider another method of solution of the same task and begin with the above representation of the generalized Cartan form $\theta$. Due to the obvious indices symmetry reasons it can be recast in 'apparently covariant' form as follows:

$$\begin{aligned} \theta =& 2\sqrt{-g}g^{\alpha\beta}g^{\nu\mu}\,X_{[\alpha\mu]}\,X_{[\beta\nu]}\,\eta \\ &+ 16\sqrt{-g}g^{\alpha_1\beta}g^{\nu\mu_1}X_{[\beta\nu]}\,\delta\psi_{\alpha_1} \wedge \eta_{\mu_1} \\ &+ 6\sqrt{-g}g^{\alpha_1\mu_1}g^{\mu_2\alpha_2}\,\delta\psi_{\alpha_1} \wedge \delta\psi_{\alpha_2} \wedge \eta_{\mu_1\mu_2}. \end{aligned}$$

The exterior differential $d$ of this expression whose all the indices are contracted can be replaced by the 'covariant' differential $D$ which is constructed in accordance with index structures of object in a way usual for tensor calculus and commutes with multiplication to the metric tensor:

$$
\begin{aligned}
&d\theta = D\theta = \\
&\quad 2g^{\alpha\beta}g^{\nu\mu}\Big(X_{[\alpha\mu]}\,DX_{[\beta\nu]}\wedge(\sqrt{-g}\eta) + X_{[\beta\nu]}\,DX_{[\alpha\mu]}\wedge(\sqrt{-g}\eta) + X_{[\alpha\mu]}\,X_{[\beta\nu]}\wedge D(\sqrt{-g}\eta)\Big) \\
&\quad + 16g^{\alpha_1\beta}g^{\nu\mu_1}\Big(DX_{[\beta\nu]}\wedge\delta\psi_{\alpha_1}\wedge\big(\sqrt{-g}\eta_{\mu_1}\big) + X_{[\beta\nu]}\,D\delta\psi_{\alpha_1}\wedge(\sqrt{-g}\eta_{\mu_1}) \\
&\qquad\qquad - X_{[\beta\nu]}\,\delta\psi_{\alpha_1}\wedge D(\sqrt{-g}\eta_{\mu_1})\Big) \\
&\quad + 6g^{\alpha_1\mu_1}g^{\mu_2\alpha_2}\Big(2\,D\delta\psi_{\alpha_1}\wedge\delta\psi_{\alpha_2}\wedge(\sqrt{-g}\eta_{\mu_1\mu_2}) + \delta\psi_{\alpha_1}\wedge\delta\psi_{\alpha_2}\wedge D(\sqrt{-g}\eta_{\mu_1\mu_2})\Big).
\end{aligned}
$$

One has obviously

$$
\begin{aligned}
D(\sqrt{-g}\eta) &= 0, \\
D(\sqrt{-g}\eta_\mu) &= (\sqrt{-g})_{,\nu}dx^\nu\wedge\eta_\mu - \sqrt{-g}\Gamma^\rho_{\mu\nu}dx^\nu\wedge\eta_\rho = \Big[(\sqrt{-g})_{,\mu} - \sqrt{-g}\Gamma^\rho_{\mu\rho}\Big]\wedge\eta = 0 \\
D(\sqrt{-g}\eta_{\mu_1\mu_2}) &= (\sqrt{-g})_{,\nu}dx^\nu\wedge\eta_{\mu_1\mu_2} - \sqrt{-g}\Gamma^\rho_{\mu_1\nu}dx^\nu\wedge\eta_{\rho\mu_2} - \sqrt{-g}\Gamma^\rho_{\mu_2\nu}dx^\nu\wedge\eta_{\mu_1\rho} \\
&= \tfrac{2}{3}(\sqrt{-g})_{,[\mu_1}\eta_{\mu_2]} - \tfrac{2}{3}\sqrt{-g}\Gamma^\rho_{\mu_1[\rho}\eta_{\mu_2]} + \tfrac{2}{3}\sqrt{-g}\Gamma^\rho_{\mu_2[\rho}\eta_{\mu_1]} \\
&= \tfrac{2}{3}(\sqrt{-g})_{,[\mu_1}\eta_{\mu_2]} - \tfrac{2}{3}\sqrt{-g}\Gamma^\rho_{\rho[\mu_1}\eta_{\mu_2]} = 0.
\end{aligned}
$$

Further,

$$\delta\psi_\alpha \equiv d\psi_\alpha - \chi_{\alpha\mu}dx^\mu = D\psi_\alpha - X_{\alpha\mu}dx^\mu$$

and therefore

$$D\delta\psi_\alpha = DD\psi_\alpha - DX_{\alpha\mu}\wedge dx^\mu - X_{\alpha\mu}Ddx^\mu.$$

But

$$Ddx^\mu = ddx^\mu + \Gamma^\mu_{\rho\nu}dx^\rho\wedge dx^\nu \equiv 0$$

and

$$D\delta\psi_\alpha = DD\psi_\alpha - DX_{\alpha\mu}\wedge dx^\mu.$$

We have the following representation:

$$
\begin{aligned}
d\theta = \;& 2\sqrt{-g}g^{\alpha\beta}g^{\nu\mu}\Big(X_{[\alpha\mu]}\,DX_{[\beta\nu]} + X_{[\beta\nu]}\,DX_{[\alpha\mu]}\Big)\wedge\eta \\
& + 16\sqrt{-g}g^{\alpha_1\beta}g^{\nu\mu_1}\Big(DX_{[\beta\nu]}\wedge\delta\psi_{\alpha_1} + X_{[\beta\nu]}\,D\delta\psi_{\alpha_1}\Big)\wedge\eta_{\mu_1} \\
& + 12\sqrt{-g}g^{\alpha_1\mu_1}g^{\mu_2\alpha_2}\,D\delta\psi_{\alpha_1}\wedge\delta\psi_{\alpha_2}\wedge\eta_{\mu_1\mu_2}.
\end{aligned}
$$

It contains the terms which are not ‘apparently covariant’ and have to be expanded:

$$D\delta\psi_{\alpha_1}\wedge\eta_{\mu_1} = DD\psi_{\alpha_1}\wedge\eta_{\mu_1} - \tfrac{1}{4}DX_{\alpha_1\mu_1}\wedge\eta,$$

$$
\begin{aligned}
&D\delta\psi_{\alpha_1}\wedge\delta\psi_{\alpha_2}\wedge\eta_{\mu_1\mu_2} \\
&\quad = DD\psi_{\alpha_1}\wedge\delta\psi_{\alpha_2}\wedge\eta_{\mu_1\mu_2} + DX_{\alpha\mu_3}\wedge\delta\psi_{\alpha_2}\wedge dx^{\mu_3}\wedge\eta_{\mu_1\mu_2} \\
&\quad = DD\psi_{\alpha_1}\wedge\delta\psi_{\alpha_2}\wedge\eta_{\mu_1\mu_2} + \tfrac{1}{3}DX_{\alpha_1\mu_1}\wedge\delta\psi_{\alpha_2}\wedge\eta_{\mu_2} - \tfrac{1}{3}DX_{\alpha_1\mu_2}\wedge\delta\psi_{\alpha_2}\wedge\eta_{\mu_1}.
\end{aligned}
$$

This yields

$$
\begin{aligned}
d\theta = & \ \underline{4\sqrt{-g}g^{\alpha\beta}g^{\nu\mu}X_{[\beta\nu]}\,DX_{[\alpha\mu]}\wedge\eta} \qquad \text{from the first line}\\
& - \underline{4\sqrt{-g}g^{\alpha\beta}g^{\nu\mu}X_{[\beta\nu]}DX_{\alpha\mu}\wedge\eta} \qquad \text{from the second line}\\
& + 16\sqrt{-g}g^{\alpha\beta}g^{\nu\mu}DX_{[\beta\nu]}\wedge\delta\psi_\alpha\wedge\eta_\mu \qquad \text{from the second line}\\
& + 4\sqrt{-g}g^{\alpha_1\mu_1}g^{\mu_2\alpha_2}\,DX_{\alpha_1\mu_1}\wedge\delta\psi_{\alpha_2}\wedge\eta_{\mu_2} \qquad \text{from the third line}\\
& - 4\sqrt{-g}g^{\alpha_1\mu_1}g^{\mu_2\alpha_2}\,DX_{\alpha_1\mu_2}\wedge\delta\psi_{\alpha_2}\wedge\eta_{\mu_1} \qquad \text{from the third line}\\
& + 16\sqrt{-g}g^{\alpha\beta}g^{\nu\mu}X_{[\beta\nu]}DD\psi_\alpha\wedge\eta_\mu \qquad \text{from the second line}\\
& + 12\sqrt{-g}g^{\alpha_1\mu_1}g^{\mu_2\alpha_2}\,DD\psi_{\alpha_1}\wedge\delta\psi_{\alpha_2}\wedge\eta_{\mu_1\mu_2} \qquad \text{from the third line}\\
= & \ 16\sqrt{-g}g^{\alpha\beta}g^{\nu\mu}DX_{[\beta\nu]}\wedge\delta\psi_\alpha\wedge\eta_\mu\\
& + 8\sqrt{-g}g^{\beta[\nu}g^{\mu]\alpha}\,DX_{\beta\nu}\wedge\delta\psi_\alpha\wedge\eta_\mu\\
& + 16\sqrt{-g}g^{\alpha\beta}g^{\nu\mu}X_{[\beta\nu]}DD\psi_\alpha\wedge\eta_\mu\\
& + 12\sqrt{-g}g^{\alpha_1\mu_1}g^{\mu_2\alpha_2}\,DD\psi_{\alpha_1}\wedge\delta\psi_{\alpha_2}\wedge\eta_{\mu_1\mu_2}.
\end{aligned}
$$

The duplicated covariant differentials can be expressed in terms of Riemann tensor as follows

$$
\begin{aligned}
DD\psi_\alpha = & dD\psi_\alpha - \Gamma^\beta_{\alpha\mu}dx^\mu\wedge D\psi_\beta\\
= & d\Big\{d\psi_\alpha - \psi_\beta\Gamma^\beta_{\alpha\mu}dx^\mu\Big\} - \Gamma^\beta_{\alpha\mu}dx^\mu\wedge\Big\{d\psi_\beta - \psi_\gamma\Gamma^\gamma_{\beta\rho}dx^\rho\Big\}\\
= & -\psi_\beta d\Gamma^\beta_{\alpha\mu}\wedge dx^\mu - \underline{\Gamma^\beta_{\alpha\mu}d\psi_\beta\wedge dx^\mu} - \underline{\Gamma^\beta_{\alpha\mu}dx^\mu\wedge d\psi_\beta} + \psi_\gamma\Gamma^\beta_{\alpha\mu}\Gamma^\gamma_{\beta\rho}dx^\mu\wedge dx^\rho\\
= & -\psi_\beta\Big\{d\Gamma^\beta_{\alpha\mu} + \Gamma^\gamma_{\alpha\mu}\Gamma^\beta_{\gamma\rho}dx^\rho\Big\}\wedge dx^\mu = -\psi_\beta\Big\{\partial_{[\rho}\Gamma^\beta_{\ \mu]\alpha} + \Gamma^\beta_{\gamma[\rho}\Gamma^\gamma_{\ \mu]\alpha}\Big\}dx^\rho\wedge dx^\mu\\
= & -\tfrac{1}{2}\psi_\beta R^\beta_{\alpha\rho\mu}dx^\rho\wedge dx^\mu
\end{aligned}
$$

and since

$$
dx^\rho\wedge dx^\mu\wedge\eta_\nu\equiv 0, \qquad dx^\rho\wedge dx^\mu\wedge\eta_{\mu_1\mu_2} = \tfrac{1}{6}\delta^\rho_{[\mu_1}\delta^\mu_{\ \mu_2]}\eta
$$

$$
\begin{aligned}
d\theta = & \ 16\sqrt{-g}g^{\alpha\beta}g^{\nu\mu}DX_{[\beta\nu]}\wedge\delta\psi_\alpha\wedge\eta_\mu\\
& + 8\sqrt{-g}g^{\beta[\nu}g^{\mu]\alpha}\,DX_{\beta\nu}\wedge\delta\psi_\alpha\wedge\eta_\mu\\
& - 4\sqrt{-g}g^{\alpha_1\mu_1}g^{\mu_2\alpha_2}\,\psi_\beta R^\beta_{\alpha_1\mu_1\mu_2}\delta\psi_{\alpha_2}\wedge\eta\\
= & \ 16\sqrt{-g}g^{\alpha\beta}g^{\nu\mu}DX_{[\beta\nu]}\wedge\delta\psi_\alpha\wedge\eta_\mu\\
& + 8\sqrt{-g}g^{\beta[\nu}g^{\mu]\alpha}\,DX_{\beta\nu}\wedge\delta\psi_\alpha\wedge\eta_\mu\\
& + \sqrt{-g}R^{\beta\alpha}\psi_\beta\delta\psi_\alpha\wedge\eta
\end{aligned}
$$

that coincides with previously obtained representation.

What comments would be relevant to the calculations exposed above? One may suggest a number of them and all of a negative sort.

At first, the replacing of the potential 1-form $\boldsymbol{\mathcal{A}} = \psi_\alpha dx^\alpha$ being the invariant geometric object in favor of the indexed components $\psi_\alpha, \chi_{\alpha\mu}$ seems to be disadvantage badly conformed with the current tendency to work in invariant notations. Besides, this makes much more complicated the index structure of the expressions under consideration.

The second disadvantage related to the former one is a too involved index structure of equations comparatively with the problem considered. It is of course more technical than principal feature but in practice it may became a serious obstruction on the way of application of method to real and much more involved than this example problems. From the point of view of computational convenience the methods seems to be not attractive. In particular, it seems doubtful to recommend implementation of the method to the real computer algebra systems. In fact, if such a system is assumed to be able to work out with arbitrary Lagrangians it should be equipped with “absolute” capabilities to process numerous contractions of indices, taking into account and properly exploiting any type of relevant symmetries *etc.*

Next, it seems rather unattractive that the basic object $\sigma$ includes effectively the terms that are related to *symmetric* part of the partial derivatives of E-M potential. Perhaps, it is connected with some more deep physical background, but there is no clear evidence of this yet.

The last relevant remark is the following: one can see that a large part of calculations leading to the Lagrange-Souriau form $\sigma$ is in fact useless since after ‘contractions’ and ‘restrictions to space-time manifold’ a number of terms required some efforts to calculate them are merely dropped.

One has to emphasize here however that the methods of descriptions of physical systems in term of evolution spaces and generalizations of Cartan form are destined mostly to study their *symmetries* rather than to derive evolution equations [3-6]. Thus it is reasonable to outline how the symmetries of E-M field considered as dynamical system are described in terms of Lagrange-Souriau form formalism.

The ‘apparently covariant’ representation of $\sigma$ obtained above immediately implies its invariance with respect to *coordinate* transformations $x^\mu \to \tilde{x}^\mu = \boldsymbol{x}^\mu(x)$, *provided* that $\psi_\alpha, \chi_{\alpha\mu}$ are transformed by the tensorial rules in accordance with standard interpretations of their indices. It is not necessary however to restrict the class of admissible transformations of $\psi_\alpha, \chi_{\alpha\mu}$ in such a way and much more general transformations may be involved in the play. Then more general group of symmetries of the system might be found. We shall consider here less general problem and restrict ourselves to the case when the *coordinate* part of transformation is trivial, i.e. $\boldsymbol{x}^\mu(x) = x^\mu$. We are guaranteed that such symmetry subgroup is still non-trivial since it contains the *gauge* transformation of E-M potential at least. Thus we shall consider the transformation of the Lagrange-Souriau form

$$
\begin{aligned}
\sigma = & - 4\big\{\sqrt{-g}g^{\alpha[\beta}g^{\nu]\mu}\big\}_{,\mu}\, \chi_{\beta\nu}\, \delta\psi_\alpha \wedge \eta \\
& + 8\sqrt{-g}\Big\{2\, g^{\alpha[\beta}g^{\nu]\mu} + g^{\beta[\nu}g^{\mu]\alpha}\Big\} d\chi_{\beta\nu} \wedge \delta\psi_\alpha \wedge \eta_\mu \\
& + 4\big\{\sqrt{-g}g^{\beta[\nu}g^{\mu]\alpha}\big\}_{,\nu}\, \delta\psi_\beta \wedge \delta\psi_\alpha \wedge \eta_\mu
\end{aligned}
$$

under the map $\boldsymbol{\mathcal{F}}$

$$x^\mu \to x^\mu, \quad \psi_\alpha \to \tilde{\psi}_\alpha = \psi_\alpha + \boldsymbol{\psi}_\alpha(x,\psi,\chi), \quad \chi_{\alpha\mu} \to \tilde{\chi}_{\alpha\mu} = \chi_{\alpha\mu} + \boldsymbol{\chi}_{\alpha\mu}(x,\psi,\chi).$$

for arbitrary functions $\boldsymbol{\psi}, \boldsymbol{\chi}$ The ordinary E-M gauge transformation corresponds to $\boldsymbol{\psi}_\alpha = \frac{\partial\Lambda(x)}{\partial x^\alpha}$, $\boldsymbol{\chi}_{\alpha\mu} = \frac{\partial^2\Lambda(x)}{\partial x^\alpha \partial x^\mu}$ for arbitrary function $\Lambda(x)$. It is easy to see that $\sigma$ is invariant with respect to it implying that gauge transformation is a symmetry. One has obviously

$$\begin{aligned}\dot{\boldsymbol{\mathcal{F}}}^*\sigma - \sigma = &- 4\{\sqrt{-g}g^{\alpha[\beta}g^{\nu]\mu}\}_{,\mu}\Big[\boldsymbol{\chi}_{\beta\nu}\,\boldsymbol{\delta\psi}_\alpha + \boldsymbol{\chi}_{\beta\nu}\,\delta\psi_\alpha + \chi_{\beta\nu}\,\boldsymbol{\delta\psi}_\alpha\Big]\wedge\eta \\ &+ 8\sqrt{-g}\Big\{2\,g^{\alpha[\beta}g^{\nu]\mu} + g^{\beta[\nu}g^{\mu]\alpha}\Big\}\Big[d\boldsymbol{\chi}_{\beta\nu}\wedge\boldsymbol{\delta\psi}_\alpha + d\chi_{\beta\nu}\wedge\boldsymbol{\delta\psi}_\alpha + d\boldsymbol{\chi}_{\beta\nu}\wedge\delta\psi_\alpha\Big]\wedge\eta_\mu \\ &+ 4\{\sqrt{-g}g^{\beta[\nu}g^{\mu]\alpha}\}_{,\nu}\Big[\boldsymbol{\delta\psi}_\beta + 2\,\delta\psi_\beta\Big]\wedge\boldsymbol{\delta\psi}_\alpha\wedge\eta_\mu\end{aligned}$$

where the evidently generalized notation is used: $\boldsymbol{\delta\psi}_\alpha \equiv d\boldsymbol{\psi}_\alpha - \boldsymbol{\chi}_{\alpha\mu}dx^\mu$. The components $\boldsymbol{\psi}, \boldsymbol{\chi}$ have to be interpreted here as functions of $x, \psi, \chi$.

Now it must be noted that *Noethern symmetries* are defined in [1,2] just as diffeomorphisms of evolution space leaving $\sigma$ unaltered, *i.e.* as solutions $\boldsymbol{\mathcal{F}}$ of the equation

$$\dot{\boldsymbol{\mathcal{F}}}^*\sigma = \sigma.$$

Strictly speaking, this restriction is too strong and it would be sufficient for the form $\dot{\boldsymbol{\mathcal{F}}}^*\sigma - \sigma$ to vanish after contraction with the lift of arbitrary vector field being restricted to the image of space-time under the immersion $\Phi$ and on solutions of the Euler-Lagrange equations only in order to map a solution to another solution that is required from a symmetry map. On the other hand, if $\sigma$ is invariant itself this implies, at list locally, the relation $\dot{\boldsymbol{\mathcal{F}}}^*\theta - \theta = d\Omega$ for some $\Omega$ and then the Noethern current can be easily constructed. Thus even if we adopt the above more restrictive definition of the symmetry, the corresponding equations *in explicit form* will be rather complicated *nonlinear* system of partial derivatives equations. Its only obvious general property is *complete integrability* (both $\dot{\boldsymbol{\mathcal{F}}}^*\sigma$ and $\sigma$ are automatically closed) that means here that no additional information can be obtained with the help of exterior differentiation of the equation. Of course, when the problem is considered in the whole generality and arbitrary diffeomorphisms of the space-time $x^\mu \to \tilde{x}^\mu = \boldsymbol{x}^\mu(x)$, are admitted the equations become much more involved. Certainly, one may obtain some their partial solutions but to describe the whole set of symmetries seems to be unrealistic goal, at least if the above form of the equations is used.

In conclusion it is worthwhile to notice that the point of view followed in these notes concerns mostly with the computational efficiency of the Lagrange-Souriau form method. It is quite possible that its other kind features would reveal some important advantages of the method as well.

**Acknowledgment**. I am grateful to the *Graduiertenkolleg* "Scientific computing" (*Köln – St. Augustin, Nordrhein-Westfalen*) for financial support and to the Institute for Theoretical Physics of the University of Cologne for hospitality.

### Appendix

The shortest way from Lagrangian to Maxwell equations seems to be the following: we may define $L[\boldsymbol{\mathcal{A}}] = \frac{1}{2}(d\boldsymbol{\mathcal{A}}) \wedge *d\boldsymbol{\mathcal{A}}$ where $*$ is Hodge dualizing and then

$$\begin{aligned}
\delta L[\boldsymbol{\mathcal{A}}, \delta\boldsymbol{\mathcal{A}}] \equiv \frac{dL[\boldsymbol{\mathcal{A}} + \varepsilon\delta\boldsymbol{\mathcal{A}}]}{d\varepsilon}_{|\varepsilon=0} &= \tfrac{1}{2}(d\delta\boldsymbol{\mathcal{A}}) \wedge *d\boldsymbol{\mathcal{A}} + \tfrac{1}{2}(d\boldsymbol{\mathcal{A}}) \wedge *d\delta\boldsymbol{\mathcal{A}} \\
&= \tfrac{1}{2}\delta\boldsymbol{\mathcal{A}} \wedge d * d\boldsymbol{\mathcal{A}} + \tfrac{1}{2}d(\delta\boldsymbol{\mathcal{A}} \wedge *d\boldsymbol{\mathcal{A}}) + \tfrac{1}{2}(*d\boldsymbol{\mathcal{A}}) \wedge d\delta\boldsymbol{\mathcal{A}} \\
&= \delta\boldsymbol{\mathcal{A}} \wedge d * d\boldsymbol{\mathcal{A}} + d(\delta\boldsymbol{\mathcal{A}} \wedge *d\boldsymbol{\mathcal{A}}).
\end{aligned}$$

The factor $d * d\boldsymbol{\mathcal{A}}$ coupled with $\delta\boldsymbol{\mathcal{A}}$ in the first term is the left side of Maxwell equations, $\delta\boldsymbol{\mathcal{A}} \wedge *d\boldsymbol{\mathcal{A}}$ is the boundary term.